\documentclass[10pt]{article}
\usepackage{osa2}
\usepackage{overcite}  %% PRODUCES SUPERSCRIPT REFERENCE CITATIONS

\begin{document}
\title{SENSITIVITY OF AN IMAGING SPACE INFRARED INTERFEROMETER}

\author{Tadashi Nakajima}
\address{National Astronomical Observatory, 2-21-1 Osawa, Mitaka,
181-8588, Japan}
\email{tadashi.nakajima@nao.ac.jp}

\centerline{and}

\author{Hideo Matsuhara}
\address{The Institute of Space and Astronautical Science,
3-1-1 Yoshinodai, Sagamihara, 229-8510, Japan}
\email{maruma@ir.isas.ac.jp}

\begin{abstract}
We study the sensitivities of space infrared interferometers.
We formulate the signal-to-noise ratios of infrared images
obtained by aperture synthesis in the presence of source
shot noise, background shot noise and detector read noise.
We consider the case in which $n$ beams are pairwise combined
at $n(n-1)/2$ detectors, and the case in which all the $n$ beams
are combined at a single detector. We apply the results to
future missions, Terrestrial Planet Finder and Darwin. We also discuss
the potential of a far-infrared interferometer for a deep
galaxy survey.
\end{abstract}

\vskip 5mm
Accepted for publication in Applied Optics
\ocis{110.3080, 110.4280, 120.3180}

\section{Introduction}
Early in the 21st century space infrared observatories
such as ASTRO-F\cite{astro-f}, Space Infrared Telescope
Facility (SIRTF)\cite{sirtf}, 
Far Infrared and Submillimeter Telescope (FIRST)\cite{first}, 
and Space Infrared Telescope for
Cosmology and Astrophysics (SPICA)\cite{nak98} 
are planned to study the
formation of planets, stars, and galaxies. The improvement
in sensitivity is so great that the source detection
limits of
these space observatories are expected to be set by source confusion,
especially in the far infrared.
To further improve the source detection limits, individual
sources need to be resolved by an infrared interferometer.
Therefore it is important to quantify the sensitivity
of a space infrared interferometer for the purpose of synthesis
imaging.

Following the observatories,
space infrared interferometers
such as the Terrestrial Planet Finder (TPF)\cite{tpf} and  
Darwin\cite{darwin}
are considered,  with an emphasis on the detection of
terrestrial planets around nearby stars.
TPF and Darwin will consist of several radiation
cooled apertures and will be capable of 
synthesis imaging as well as planet detection by nulling
interferometry. The detection limits of these interferometers
for general synthesis imaging are of great interest.
Among many applications of  synthesis imaging
with a space infrared interferometer, 
we are  particularly interested in deep galaxy count
by which the history of galaxy formation can be studied.
We discuss how 
a TPF-like interferometer operated in the
far infrared can improve the number count data
of high redshift galaxies.

The beam combination geometry is a major issue in studying
the sensitivity of an interferometer.
One extreme is the $^nC_2$ interferometer, in which $n$ beams
are divided into $n(n-1)$ subbeams and they are pairwise
combined at $^nC_2 = n(n-1)/2$ detectors. There is
one detector for each baseline. The other extreme is the $^nC_n$
interferometer in which all the $n$ beams are combined
at $^nC_n = 1$ detector.
Prasad and Kulkarni\cite{pra89} (henceforth PK89)
have rigorously analyzed the SNRs of the $^nC_2$ and
$^nC_n$ interferometers for spectrally narrow  light in
the presence of source shot noise  and 
concluded that the sensitivities of both interferometers are
essentially the same apart from a numerical factor.
The results by PK89 are applicable to optical interferometry,
but not to infrared interferometry because of the presence
of thermal background and detector read noise.
The beam combination geometry can have an impact on the 
sensitivity in the infrared for instance because of
the number of detector pixels which affects the total read noise.
The beam combination geometry
may also have a major impact on the architecture of the space interferometer.

Baseline redundancy is a factor that affects the SNR of
a synthesis image for a given number of apertures.
Since the number of apertures deployed or flown separately
in space is likely to be limited, it is important to gain
$uv$ coverage by using nonredundant baselines. Throughout this 
paper, we only consider interferometers with nonredundant
baselines.

There are two types of noise which appear in aperture
synthesis, one of which is removable while the other is
intrinsic and not removable.
Sampling noise is potentially
introduced by the incompleteness of $uv$ coverage.
However, deconvolution techniques
such as CLEAN and Maximum Entropy Method (MEM)\cite{per89},
when applied to radio interferometric data,
appear to compensate for sampling noise.
We thus regard sampling noise as removable noise.
On the other hand, noise resulting from the photodetection process
is not removable.
The SNR of a dirty image to which deconvolution has not been
applied is limited by photodetection noise.
Here we derive the SNR of a dirty image as determined by source shot noise,
thermal background shot noise, and detector read noise,
and equate it to the SNR of the synthesized image. 

There are two different methods by which a synthesized
image can be constructed from the visibility data: inversion
without total counts and true inversion.
In the former, the zero frequency phasor is neglected, 
implying that the total number of photons in the
synthesized image is zero. 
Despite this seemingly unattractive feature, this is the
standard method in radio astronomy.
In the latter method, the van Cittert-Zernike
theorem is strictly applied and the zero frequency
phasor as well as the $n(n-1)/2$ complex
phasors is used. The image produced by this technique has the
desirable property of nonnegativity.
In the presence of additive thermal background which overwhelms
the signal, the magnitude of the zero frequency phasor is much greater than
those of other phasors.  
The zero frequency phasor determines the DC offset level due to
thermal background in the synthesized image. Not only is this
background of little astronomical interest, it is also a major source
of Poisson fluctuations. Therefore
in this paper we prefer the former method  and consider only
the inversion without total counts as in radio astronomy.

We analyze the SNRs of space interferometers in three steps.
First,
we formulate the SNRs of $^nC_2$ and
$^nC_n$ interferometers  in the presence of
source shot noise, thermal background shot noise and
detector read noise in \S2 and \S3 respectively.
These formulae cover optical and infrared
interferometers. Second, we consider spectrally broad light for which
fringe dispersion becomes an issue.
Fringes of the $^nC_2$ interferometer can always be dispersed while
those of the $^nC_n$ interferometer can be dispersed only in the case
of a
one dimensional baseline configuration.
We discuss the tradeoff between the $^nC_2$ and $^nC_n$
interferometers including the dimensionality of the baseline
configuration in \S4. 
Third, we apply the results of our analysis to
planned interferometers. 
We present the sensitivities of 
a TPF-like interferometer and a Darwin-like
one in \S5. Finally in \S6 we discuss the capability
of the TPF-like interferometer from the point of view of
deep galaxy count.

\section{$^nC_2$ Interferometer}

\subsection{Fringe Phasor Estimator}

Let there be $n$ identical principal apertures from
which we derive $n$ main beams. Each beam is divided into
$n-1$ subbeams by the use of beam splitters. The resulting
$n(n-1)$ subbeams are combined pairwise on $n_b = n(n-1)/2$
detectors where $n_b$ is also the number of baselines.

The fringe pattern is formed at the focal plane of
each detector. For spectrally narrow light,
a one-dimensional detector is sufficient (Fig. 1). 
For spectrally broad light, the fringes can be dispersed
in the cross fringe direction, in which case a two-dimensional
detector is used. 
For a focal plane interferometer of this type,
the influence of source shot noise on 
fringe phasor estimation has
been fully formulated\cite{wal73}$^,$\cite{goo85}. 
Let the interferometer be illuminated by a source upon a
spatially smooth background.
The intensity pattern on the  $r^{\rm th}$ detector ($r^{\rm th}$ baseline)
is given by

\begin{equation}
<I_r({\bf x})> = 2<I_0> \Bigl[1+\gamma_r
                 \cos(\kappa{\bf x}\cdot {\bf B}_r/d + \phi_r)\Bigr],
\end{equation}

where $<I_0>$ is the average intensity in each subbeam at the
detector, ${\bf B}_r$ is the baseline vector, 
$\kappa=2\pi/\lambda$ is the wave number, and $d$ is 
the distance between the aperture plane and the detector,
${\bf x}$ is the spatial vector in the detector plane,
and $\gamma_r \exp(i\phi_r)$ is the complex visibility function
at the baseline vector ${\bf B}_r$. In deriving (1), we have
assumed that the incident light is spectrally narrow so that
the fringe visibility depends only on the spatial correlation of
the field. 

%\placefigure{fig1}

Because of the presence of background illumination,

\begin{equation}
I_0 = I_0^s + I_0^b,
\end{equation}

where $I_0^s$ and $I_0^b$ are source and background intensities
respectively. Let $\gamma^s_r$ be the fringe visibility of the source
in the absence of the background, then

\begin{equation}
\gamma_r = \frac{I_0^s}{I_0^s + I_0^b} \gamma^s_r.
\end{equation}

The complex visibility function $\gamma^s_r \exp(i\phi_r)$
is determined by the $uv$ coordinates of the $r^{\rm th}$ baseline $(u,v)$ 
and the source
brightness distribution on the sky $S(x,y)$ by a Fourier transform relation:

\begin{equation}
\gamma^s_r \exp(i\phi_r)
=\frac{\int S(x,y)\exp\{-2 \pi i(ux+vy)\} dx dy}
      {\int S(x,y) dx dy}.
\end{equation}

In an effort to reduce the clutter in the equations we henceforth
drop the vector notation, but bear in mind that spatial frequencies,
pixel locations, etc. are really vectors.
The photoelectron detection theory\cite{wal73} 
takes into account the discrete
nature of both photons and detector pixels.  
The average 
photoelectron count $<k_r(p)>$ at the pixel location specified by
the integer index $p$ of the detector is proportional to
$<I(x)>$: 
 
\begin{equation}
<k_r(p)> = 2<K_0>\Bigl[1+\gamma_r\cos(p\omega_r + \phi_r)\Bigr].
\end{equation}

Here, $<...>$ denotes averaging over the photoelectron-detection
process. 
$K_0$ is the discrete version of $I_0$.
The product $p\omega_r$ is understood to be the scalar
product of the pixel position vector ${\bf p}$ and the spatial
frequency ${\bf \omega}_r$ expressed in inverse pixel units.

Let $<C>$ be the average number of photoelectrons detected by
the entire interferometer in one integration period, and let 2$<N>$ be
the average number of photoelectrons per detector of
each baseline per integration
time. Clearly then, $<C> = 2<N>n_b$, and thus $<N> = 
<C>/\{n(n-1)\}$.  According to (5), the average number of
photoelectrons per detector is equal to $2<K_0>P$, and thus
$<K_0>P = <N>$, where $P$ is the number of detector pixels.

Due to read noise $r_r(p)$,  each measurement of the pixel $p$
yields $l_r(p)$ given by

\begin{equation}
l_r(p) = k_r(p) + r_r(p),
\end{equation}

where $r_r(p)$ is a Gaussian random number with zero mean and
a standard deviation $\sigma$. 

The fringe phasor for the baseline $r$ is operationally defined
as

\begin{equation}
y_r = \sum_{p=1}^P l_r(p) \exp(-ip\omega_r).
\end{equation}

Walkup and Goodman  have shown that the quantity

\begin{equation}
z_r = \sum_{p=1}^P k_r(p) \exp(-ip\omega_r)
\end{equation}

is an optimum estimator of the actual fringe phasor 
under shot noise limited condition in the presence of
constant background\cite{wal73}. Since 

\begin{equation}
<\sum_{p=1}^P r_r(p) \exp(-ip\omega_r)> 
=\sum_{p=1}^P <r_r(p)> \exp(-ip\omega_r) = 0,
\end{equation}

$y_r$ is an optimum estimator of the fringe phasor
in the presence of detector read noise. 

The average of the phasor $y_r$ over many coherent
integration interval is given by

\begin{equation}
Y_r = <y_r> = <N> \gamma_r \exp(i\phi_r).
\end{equation}

\subsection{Inversion}

The synthesized image is the real portion of the Fourier
transform of the spatial coherence function:

\begin{eqnarray}
i_1(q) & = & Re \Bigl[\sum_{r=1}^{n_b} y_r \exp(+i\omega_r q)\Bigr]
\nonumber \\
& = & \sum_r [Re(y_r)\cos(\omega_r q) - Im(y_r) \sin(\omega_r q)].
\end{eqnarray}

Index $q$ refers to pixels in the synthesized image; in particular
$q$ ranges from $-Q/2$ to $+Q/2$, and $q=0$ refers to the central
pixel in the synthesized image. 
The variable $i_1(q)$ refers to the image obtained from one
set of visibility data. Note that the sense of Fourier transform
used in (11) is consistent with the definition of $y_r$.
The mean image $I_1(q)$ is the average of $i_1(q)$ and is given by
 
\begin{equation}
I_1(q) = Re \Bigr[\sum_{r=1}^{n_b} Y_r \exp(+i\omega_r q)\Bigl],
\end{equation}

which by virtue of (10), can be simplified to yield

\begin{eqnarray}
I_1(q) & = & <N>\sum_r \gamma_r \cos(\omega_r q + \phi_r) \nonumber \\
& = & <N>\sum_r \frac{I^s_0}{I^s_0+I^b_0}\gamma_r^s \cos(\omega_r q + \phi_r).
\end{eqnarray}

The image $I_1(q)$ is referred to as the dirty image 
in the parlance of radio astronomy. 
The dirty image is the convolution of the true image
and the Fourier transform of the spatial frequency sampling
function or the dirty beam. A synthesized image can be
obtained from the dirty image by deconvolution.

The variance $V[i_1(q)]$ in the synthesized image $i_1(q)$
is given by

\begin{eqnarray}
V[i_1(q)] & = & <i_1(q)^2> - <i_1(q)>^2 \nonumber \\
& = & \sum_{r=1}^{n_b}\sum_{s=1}^{n_b} \Bigr\{ 
[<Re(y_r)Re(y_s)> - <Re(y_r)><Re(y_s)>]\cos(\omega_rq)\cos(\omega_sq) 
\nonumber \\
& & 
-[<Re(y_r)Im(y_s)> - <Re(y_r)><Im(y_s)>]\cos(\omega_rq)\sin(\omega_sq)
\nonumber \\
& &
-[<Im(y_r)Re(y_s)> - <Im(y_r)><Re(y_s)>]\sin(\omega_rq)\cos(\omega_sq)
\nonumber \\
& &
+[<Im(y_r)Im(y_s)> - <Im(y_r)><Im(y_s)>]\sin(\omega_rq)\sin(\omega_sq)
\Bigl\}.
\end{eqnarray}

The problem of estimating the variance of the image reduces to that
of estimating three types of covariance term: 
$cov[Re(y_r),Re(y_s)]$, $cov[Re(y_r),Im(y_s)]$, 
and $cov[Im(y_r),Im(y_s)]$.

\begin{eqnarray}
cov[Re(y_r),Re(y_s)] &=& [<Re(y_r)Re(y_s)> - <Re(y_r)><Re(y_s)>]
\nonumber \\
& = & \sum_{p=1}^P\sum_{p^\prime=1}^P
[<l_r(p)l_s(p^\prime)>-<l_r(p)><l_s(p^\prime)>]
\cos(\omega_rp)\cos(\omega_rp^\prime).
\end{eqnarray}

where

\begin{eqnarray}
 <l_r(p)l_s(p^\prime)>-<l_r(p)><l_s(p^\prime)> 
& = & <k_r(p)k_s(p^\prime)>-<k_r(p)><k_s(p^\prime)> \nonumber \\ 
&  &     +<k_r(p)r_s(p^\prime)>-<k_r(p)><r_s(p^\prime)> \nonumber \\ 
& &     +<r_r(p)k_s(p^\prime)>-<r_r(p)><k_s(p^\prime)> \nonumber \\ 
& &     +<r_r(p)r_s(p^\prime)>-<r_r(p)><r_s(p^\prime)>.
\end{eqnarray}

There is no correlation of shot noise between different detectors
or between different pixels of the same detector.
The correlation remains only for the same pixel. 
There is no correlation of shot noise and read noise even for
the same detector pixel. And there is no correlation of read noise
between different detectors or between different pixels of the same
detector. The correlation exists only for the same pixel.
Therefore

\begin{equation}
<k_r(p)k_s(p^\prime)>-<k_r(p)><k_s(p^\prime)>
=\delta_{rs}\delta_{pp^\prime}<k_r(p)>,
\end{equation}

\begin{equation}
<k_r(p)r_s(p^\prime)>-<k_r(p)><r_s(p^\prime)>=
<r_r(p)k_s(p^\prime)>-<r_r(p)><k_s(p^\prime)>=0,
\end{equation}

and 

\begin{equation}
<r_r(p)r_s(p^\prime)>-<r_r(p)><r_s(p^\prime)>
=\delta_{rs}\delta_{pp^\prime}\sigma^2.
\end{equation}

Therefore

\begin{equation}
<l_r(p)l_s(p^\prime)>-<l_r(p)><l_s(p^\prime)>
=\delta_{rs}\delta_{pp^\prime}(<k_r(p)>+\sigma^2),
\end{equation}

and 

\begin{eqnarray}
cov[Re(y_r),Re(y_s)] 
& = & V[Re(y_r)]\delta_{rs} \nonumber \\
& = & \delta_{rs}\sum_p(<k_r(p)>+\sigma^2)\cos(\omega_r p)^2
\nonumber \\
& = & \delta_{rs} (<N> + \frac{P\sigma^2}{2}).
\end{eqnarray}

Similarly, one can show that

\begin{equation}
cov[Re(y_r),Im(y_s)] = cov[Im(y_r),Re(y_s)] = 0,
\end{equation}

and

\begin{equation}
cov[Im(y_r),Im(y_s)] = \delta_{rs} (<N> + \frac{P\sigma^2}{2}). 
\end{equation}

Substituting the nonzero covariance elements (21) and (23) into (14),
we obtain

\begin{eqnarray}
V[i_1(q)] & = & 
\sum_r\Bigr\{(<N>+\frac{P\sigma^2}{2}) \cos^2(\omega_rq)
+(<N>+\frac{P\sigma^2}{2}) \sin^2(\omega_rq)\Bigl\}
\nonumber \\
& = & \sum_r (<N>+\frac{P\sigma^2}{2}) \nonumber \\
& = & n_b (<N>+\frac{P\sigma^2}{2}) 
  = \frac{<C>}{2} + n_b \frac{P\sigma^2}{2}.
\end{eqnarray}

In (24), the term $<C>/2$ is due to shot noise and the term 
$n_b P \sigma^2 / 2$ is due to read noise.
Now we consider the specific case of a point source
($\gamma^s_r = 1$) at the phase center ($\phi_r = 0$) for which
$I_1(0) = \frac{<C>}{2}$.
Since the source is located at the phase center, the SNR of
the central pixel in the image is indicative of the SNR in the image:

\begin{equation}
\frac{I_1(0)}{\sqrt{V[i_1(0)]}}
=\frac{ (<C>/2) \{ I_0^s/(I_0^s+I_0^b)\} }
 {\sqrt{ <C>/2 + n_bP\sigma^2/2 }}.
\end{equation}

\section{$^n C_n$ Interferometer}

In an $^nC_n$ interferometer, all  $n$ beams interfere
on a single detector and $n_b$ fringes are superposed.
Both the baseline configuration and the detector can be
either one dimensional or two dimensional. 
As an example, a two dimensional 
$^3C_3$ interferometer is schematically presented
in Fig. 2. 
A special case
is the combination of a one dimensional baseline configuration
and a two dimensional detector for which the superposed fringes
are dispersed in the cross fringe direction so that
spectrally broad light can be used without bandwidth smearing.
We call an $^nC_n$ interferometer of this type an  $^nC_n^\prime$ 
interferometer.  A $^3C_3^\prime$ interferometer is schematically
shown in Fig. 3.
Application of (7) with different spatial frequencies results
in extraction of $n_b$ fringe phasors and the image may be
synthesized in the usual fashion.

%\placefigure{fig2}

%\placefigure{fig3}

\subsection{Fringe phasor estimator}

Let the interferometer be composed of $n$ identical apertures
and let it be illuminated by a source and a spatially smooth
background.
The classical intensity distribution of the interference
pattern for the $n$ apertures has the average value

\begin{equation}
<I({\bf x})> = <I_0> \Bigl[ n + 2 \sum_{g<h} \gamma_{gh}
\cos(\kappa{\bf x}\cdot {\bf B}_{gh}/d + \phi_{gh}) \Bigr],
\end{equation}

where the various symbols have meanings similar to those in (1).
$gh$  denotes the baseline $gh$ corresponding to the apertures 
$g$ and $h$. Let $<k({\bf p})>$ denote the photoelectron count
distribution due to $<I({\bf x})>$. As in \S2, we discontinue
the vector notation, assume that the total number of pixels is $P$,
and note that $<k(p)>$ is proportional to $<I(x)>$:

\begin{equation}
<k(p)> = <Q_0> \Bigl[n + \sum_{g<h}\gamma_{gh}
\cos(p\omega_{gh} + \phi_{gh}) \Bigr].
\end{equation}

Here $<Q_0>$ has approximately the same meaning as $<K_0>$ in
\S2. However, since there is no beam splitting,
 $<Q_0> = (n-1)<K_0> $. 
Due to read noise $r(p)$, each measurement of
the pixel $p$ yields $ l(p)$ counts:

\begin{equation}
l(p) = k(p) + r(p),  
\end{equation}

where $r(p)$ is a Gaussian random number with zero mean and
a standard deviation $\sigma$.

In a typical setup, one can imagine integrating on a 
two dimensional detector for a period equal to
the coherent integration interval. A Fourier transform
of the resulting image yields 2$n_b$ peaks, pairwise
symmetrical about the origin, which can
be identified with the $n_b$ fringe phasors corresponding
to the $n_b$ spatial-frequency components.
Let $y_{gh}$ be the fringe phasor estimator
corresponding to the baseline $gh$,
then

\begin{equation}
y_{gh} = \sum_{p=1}^{P}l(p)\exp(-ip\omega_{gh}).
\end{equation}

Here $y_{gh}$ is an optimum estimator of the fringe phasor for
the baseline $gh$.
The mean phasor on the $ij$ baseline is given by

\begin{eqnarray}
Y_{ij} & = & <Q_0> \sum_p \exp(-ip\omega_{ij}) 
\Bigl[ n+2\sum_{g<h}\gamma_{gh}\cos(p\omega_{gh}+\phi_{gh})\Bigr]
+  \sum_{p}<r(p)>\exp(-ip\omega_{ij})
\nonumber \\
& = & <Q_0> \sum_p\sum_{g<h}\gamma_{gh}
\Bigl\{\exp(i\phi_{gh})\exp[ip(\omega_{gh}-\omega_{ij})] \nonumber \\
&  & \ \ \ \ \ \ \ \ \ \ \ 
 + \exp(-i\phi_{gh})\exp[-ip(\omega_{gh} + \omega_{ij})]\Bigr\}.
\end{eqnarray}

Since the baselines are nonredundant, $\omega_{ij} \neq \omega_{gh},$
unless $(ij)$ and $(gh)$ refer to the same baseline. Therefore
only the $g=i, h=j$ terms remains in (30):

\begin{equation}
Y_{ij} = Z_{ij} = <M> \gamma_{ij} \exp(i\phi_{ij}),
\end{equation}

where we define $<M> = P<Q_0> = <C>/n$.

\subsection{Inversion}

The dirty image, $i_2(q)$ is given by

\begin{eqnarray}
i_2(q) & = & Re\Bigl[\sum_{i<j} y_{ij} \exp(iq\omega_{ij})\Bigr]
\nonumber \\
& = & \sum_{i<j}
\Bigl[Re(y_{ij})\cos(q\omega_{ij})-Im(y_{ij})\sin(q\omega_{ij})\Bigr].
\end{eqnarray}

Using (31), we find
the mean synthesized image to be

\begin{eqnarray}
I_2(q) &  = & Re\Bigl[\sum_{i<j} Y_{ij} 
\exp(+iq\omega_{ij})\Bigr] \nonumber \\
       &  = & <M> \sum_{i<j} \gamma_{ij}\cos(q\omega_{ij}+\phi_{ij}).
\end{eqnarray}

The derivation of 
the variance $V[i_2(q)]$ in the synthesized image $i_2(q)$
is somewhat lengthy and given in Appendix A.
The result is:

\begin{eqnarray}
V[i_2(q)] & = & \frac{n<M>+P\sigma^2}{2}n_b+<M>(n-2)
\sum_{i<j}\gamma_{ij}\cos(q\omega_{ij}+\phi_{ij}) \nonumber \\
& = &  \frac{n<M>+P\sigma^2}{2}n_b+<M>(n-2)
\sum_{i<j}\frac{I^s_0}{I^s_0+I^b_0}\gamma^s_{ij}
\cos(q\omega_{ij}+\phi_{ij})
.
\end{eqnarray}

For a point source at the phase center,
$q = 0, \gamma^s_{ij} = 1, \phi_{ij}=0$ and the SNR of the
synthesized image is 

\begin{eqnarray}
\frac{I_1(0)}{\sqrt{V[i_1(0)]}}  & = &
\frac
{
<M> \{I_0^s/(I_0^s+I_0^b)\} \sqrt{n_b}
}
{\sqrt{ (n<M>+P\sigma^2)/2+<M>(n-2)
\{I_0^s/(I_0^s+I_0^b)\}} 
} \nonumber \\
& = &
\frac{<C>\{I_0^s/(I_0^s+I_0^b)\}\sqrt{(n-1)/n} }
 {\sqrt{<C>+P\sigma^2 + <C>\{I_0^s/(I_0^s+I_0^b)\}
     \{2(n-2)/n\} }}. 
\end{eqnarray}

\section{Tradeoff among $^nC_2$, $^nC_n$ and $^nC_n^\prime$ Interferometers}

We now compare the SNRs of the $^nC_2$, $^nC_n$ and   $^nC_n^\prime$ 
interferometers in the case  of a point source at the phase center.
We are interested in the detection limits of these three
interferometers and we consider the situation in which
$I_0^s \ll I_0^b$. The total photoelectron count $<C>$ is mostly
due to thermal background.  
In the limit of $I_0^s \ll I_0^b$,
the SNR of the $^nC_2$ interferometer, $SNR_1$, is given by

\begin{equation}
SNR_1 =\frac{ (<C_1>/2) \{ I_0^s/I_0^b \} }
 {\sqrt{ <C_1>/2 + n_bP_1\sigma^2/2 }},
\end{equation}

where $<C_1>$ is the total photoelectron count and $P_1$ is
the total number of pixels per detector. The SNR of 
the $^nC_n$ interferometer, $SNR_2$ is given by

\begin{equation}
SNR_2 = \frac{<C_2>\{I_0^s/I_0^b\}\sqrt{(n-1)/n} }
 {\sqrt{<C_2>+P_2\sigma^2}}, 
\end{equation}

where $<C_2>$ is the total photoelectron count and $P_2$ is
the total number of detector pixels. 

The SNR of the  $^nC_n^\prime$ interferometer, $SNR_2^\prime$ is given by

\begin{equation}
SNR_2^\prime = \frac{<C_2^\prime>\{I_0^s/I_0^b\}\sqrt{(n-1)/n} }
 {\sqrt{<C_2^\prime>+P_2^\prime\sigma^2}}, 
\end{equation}

where $<C_2^\prime>$ is the total photoelectron count and $P_2^\prime$ is
the total number of detector pixels on which superposed
fringes are dispersed. 

$<C_1>$, $<C_2>$, and $<C_2^\prime>$ 
depend on the spectral bandwidths,
the optical throughputs and detector quantum efficiencies
of the interferometers. Here for simplicity we assume that
the optical throughputs and detector quantum efficiencies are the same,
though in practice the throughputs 
may differ due to the difference in optical configurations.
$P_1$, $P_2$ and $P_2^\prime$ 
are determined by the fringe sampling requirements.

Expressions (25) and (35) were  derived for spectrally narrow light.
In practice this corresponds to the observations of emission lines.
For bright objects, it is always advantageous to disperse fringes
and reconstruct images at different wavelengths. Especially for
emission line objects, spectral dispersion is essential.
In this case, (36) and (38) hold for the fringe row corresponding
to each spectral bin. For faint objects with broad spectral
energy distributions, spectral images formed from individual 
dispersed fringes need to be coadded to increase detection
sensitivities. Below we confirm that 
(36) and (38) hold also for coadded images of the faint objects.

We first consider the case of the $^nC_2$ interferometer.
Let $P_x$ be the number of pixels in the fringe direction
and $P_y$ be that in the cross fringe direction.
Therefore $P_1 = P_x P_y$.
We assume that the photon spectrum within the bandwidth
is not steep and the number of photons in 
one detector row is given by $<c_1> = <C_1>/P_y$.
Then the SNR of an image synthesized from  a set of fringes
in detector rows
corresponding to $1/P_y$ of the spectral bandwidth 
is given by

 \begin{equation}
snr_1 =\frac{ (<c_1>/2) \{ I_0^s/I_0^b \} }
 {\sqrt{ <c_1>/2 + n_bP_x\sigma^2/2 }}.
\end{equation}

We further assume the total bandwidth is narrow enough so that
the source structure does not change within it.
In this case, $P_y$ images  
can be coadded to form a final image. The SNR of the final image
is given by

 \begin{eqnarray}
SNR_1 & = & \sqrt{P_y} snr_1 \nonumber \\
      & = & \sqrt{P_y} \frac{ (<c_1>/2) \{ I_0^s/I_0^b \} }
 {\sqrt{ <c_1>/2 + n_bP_x\sigma^2/2 }} \nonumber \\
      & = &\frac{ (<C_1>/2) \{ I_0^s/I_0^b \} }
 {\sqrt{ <C_1>/2 + n_bP_xP_y\sigma^2/2 }} \nonumber \\
      & = & \frac{ (<C_1>/2) \{ I_0^s/I_0^b \} }
 {\sqrt{ <C_1>/2 + n_bP_1\sigma^2/2 }}. 
\end{eqnarray}

Therefore (36) holds for spectrally broad light.
Similarly one can show that (38) holds for spectrally broad light.

\subsection{Background Limit For Spectrally Narrow Light}

For the $^nC_2$ interferometer, the condition for background
limited observations 
is $<C_1> \gg n_b P_1 \sigma^2$. In this case, the SNR is
approximated by

\begin{equation}
SNR_1 = \sqrt{\frac{<C_1>}{2}}\frac{I_0^s}{I_0^b}.
\end{equation}

For the $^nC_n$ interferometer, the condition for
background limit  is $<C_2> \gg P_2 \sigma^2$. In this case, the SNR is
approximated by

\begin{equation}
SNR_2 = \sqrt{<C_2>}\frac{I_0^s}{I_0^b}\sqrt{\frac{n-1}{n}}.
\end{equation}

For spectrally narrow light, the spectral bandwidths do not matter and
$<C_1> = <C_2>$. Then

\begin{equation}
SNR_2 = \sqrt{\frac{2(n-1)}{n}} SNR_1,
\end{equation}

Since $1< \sqrt{2(n-1)/n} < \sqrt{2}$, for $n>3$, 
the $^nC_n$ interferometer is preferred
in this case.  However they differ by at most a factor
$\sqrt{2}$ for a large $n$. There is no difference between
the  $^nC_n$ interferometer and  the  
$^nC_n^\prime$ interferometer for spectrally narrow light.

\subsection{Background Limit for Spectrally Broad Light}

The fractional bandwidth of the $^nC_2$ interferometer, 
$\Delta\nu/\nu$ is limited only by
the wavelength dependence
of the astronomical source structure. Each fringe
can be spectrally dispersed on a two dimensional detector
to avoid bandwidth smearing. In the general
$^nC_n$ interferometer with a two dimensional baseline configuration,
all the fringes are superposed on a two dimensional detector
and they cannot be dispersed. 
To keep the fringe visibilities high, the fractional bandwidth
is limited approximately to $D/B$ where $D$ is the aperture diameter
and $B$ is the length of the longest baseline.
In the $^nC_n^\prime$ interferometer with a one dimensional
baseline configuration, dispersed fringes are superposed on
a two dimensional detector and the fractional bandwidth, $\Delta\nu/\nu$,
is the same as for the $^nC_2$ interferometer.  
Therefore

\begin{equation}
C_1 = C_2^\prime,
\end{equation}

where $C_2^\prime$ is the total number of photoelectron count
for the $^nC_n^\prime$ interferometer. Since

\begin{equation}
C_1/C_2 = \frac{\Delta\nu/\nu}{D/B},
\end{equation}

the SNR ratio of the $^nC_2$ and $^nC_n$ interferometers is given by

\begin{eqnarray}
SNR_1/SNR_2 & = & 
\sqrt{ \frac{<C_1>}{<C_2>} \frac{n}{2(n-1)} }  \nonumber \\
            & = & \sqrt{\frac{\Delta\nu}{\nu}\frac{B}{D}\frac{n}{2(n-1)}}.
\end{eqnarray}

For a short baseline (small $B/D$), the $^nC_n$ interferometer is 
superior and for a long baseline, the $^nC_2$ has an advantage.
From $SNR_1 = SNR_2$, we obtain $(B/D)_{cross \ over}$ given by

\begin{equation}
(\frac{B}{D})_{cross \ over}
= \frac{\nu}{\Delta\nu}\frac{2(n-1)}{n}.
\end{equation}

For a continuum source, the fractional bandwidth 
$\Delta\nu/\nu$ for which
the source structure does not change is the order of 0.1. 
For $n=5$,  
$(B/D)_{cross \ over}$ = 16.

From (43), the ratio of SNR for the $^nC_2$ and $^nC_n^\prime$
interferometers is given by

\begin{equation}
SNR_1/SNR_2^\prime  =  
\sqrt{ \frac{n}{2(n-1)} },
\end{equation}

as in the case of spectrally narrow light.
To be fair, we should compare the same one dimensional
baseline configuration with different beam combination geometries.
In this case, the $^nC_n^\prime$ interferometer is superior to
the $^nC_2$ interferometer by at most a factor $\sqrt{2}$.

\subsection{Read noise limit}

In this subsection, we evaluate
the numbers of detector pixels,
$P_1$, $P_2$ and $P_2^\prime$.
There are $2\nu/\Delta\nu$ fringe cycles in each
interferogram. For the Nyquist sampling theorem,  
$4\nu/\Delta\nu$ pixels are needed in the fringe
direction. 
For
the $^nC_2$ and $^nC_n^\prime$ interferometers,
we consider the number of pixels in the cross fringe
direction (dispersion direction) in each dispersed fringe.
The fractional spectral bandwidth $\Delta\nu^*/\nu$
 for which bandwidth smearing
is small, is given by  $\Delta\nu^*/\nu < D/B$.
Therefore for a spectral bandwidth of $\Delta\nu$,
we need $\Delta\nu/\Delta\nu^* > (\Delta\nu/\nu)/(D/B)$
pixels in the dispersion direction. Here we take 
$2(\Delta\nu/\nu)/(D/B)$ pixels as the number of pixels
in the cross 
fringe direction. So the minimum number of pixels of the two
dimensional detector is
$P_1 = P_2^\prime = 4\nu/\Delta\nu \times 2(\Delta\nu/\nu)/(D/B) = 8(B/D)$.
For the $^nC_n$ interferometer, the longest baseline can be
aligned to one of the two sides of a two dimensional detector
and the minimum number of pixels is 
$P_2 = (4\nu/\Delta\nu)^2 = 16(B/D)^2$,
and the maximum allowed fractional bandwidth is $D/B$.

In the read noise limit,

\begin{equation}
SNR_1 = \frac{(<C_1>/2)(I_0^s/I_0^b)}
                { \sqrt{n_bP_1\sigma^2/2} },
\end{equation}

\begin{equation}
SNR_2 = \frac{ <C_2>(I_0^s/I_0^b)\sqrt{(n-1)/n} }
                { \sqrt{P_2\sigma^2/2} },
\end{equation}

and 

\begin{equation}
SNR_2^\prime = \frac{ <C_2^\prime>(I_0^s/I_0^b)\sqrt{(n-1)/n} }
                { \sqrt{P_2^\prime\sigma^2} },
\end{equation}

where (44) and (45) hold among $<C_1>$, $<C_2>$, and $<C_2^\prime>$.
For a two dimensional baseline configuration,
the SNR ratio of the $^nC_2$ and $^nC_n$ interferometers is given by

\begin{equation}
SNR_1/SNR_2 =
\frac{\sqrt{2}}{n-1}\frac{\Delta\nu}{\nu}(\frac{B}{D})^{3/2}.
\end{equation}

For a short baseline the $^nC_n$ interferometer
is superior and for a long baseline the $^nC_2$ interferometer
is better. From the condition that $SNR_1 = SNR_2$,

\begin{equation}
(\frac{B}{D})_{cross \ over}
= \Bigl[\frac{(n-1)^2}{2}(\frac{\nu}{\Delta\nu})^2\Bigr]^\frac{1}{3}.
\end{equation}

For $n=5,$ $\Delta\nu/\nu = 0.1$,
$(B/D)_{cross \ over}$ = 9.3.
For a one dimensional baseline configuration,
the SNR ratio of the $^nC_2$ and $^nC_n^\prime$ interferometers is given by

\begin{equation}
SNR_1/SNR_2^\prime =\frac{1}{n-1}.
\end{equation}

Since $SNR_1 < SNR_2^\prime$ for $n > 3$,
the $^nC_n^\prime$ interferometer is always superior to 
the $^nC_2$ interferometer in the read noise limit.

In summary, the $^nC_n$ interferometer with a two dimensional
baseline configuration has a limited use for a short baseline
interferometer or possibly a deployable space interferometer.
For a separated spacecraft interferometer with long baselines,
the $^nC_2$ is more suitable and the $^nC_n^\prime$ interferometer
with a one dimensional baseline configuration is even better.
In the background limit, the advantage of  $^nC_n^\prime$
interferometer is small, while in the read noise limit, that
is significant. This is simply because of the difference in
the number of detectors or that of detector pixels which affect
the total read noise.

\section{Sensitivity in the Presence of Natural Thermal Background}

\subsection{Natural Thermal Background}

We are most interested in the potential of an infrared interferometer
in the presence of unavoidable natural thermal background.
Here we consider the emission from interplanetary dust (IPD)
and interstellar dust (ISD) as the natural thermal background.

The brightness of IPD emission depends on ecliptic coordinates\cite{kelsall}. 
Lockman Hole ($\lambda = 137^\circ$, $\beta = 46^\circ$)
which is close to the ecliptic pole
is the representative of the dark portion of the sky
and the ecliptic plane ($\lambda = 122^\circ$, $\beta = 0^\circ$) 
is that of the bright portion of the sky apart from the 
galactic plane. 
IPD emission of the latter is brighter than that of the former
by a factor 3 to 4. The spectrum obtained by the Diffuse Infrared
Background Experiment (DIRBE) on board the Cosmic Background Explorer
(COBE)
of the IPD emission we use is
given in Ref. 11.
%The wavelength coverage of DIRBE is discrete, so we have
%interpolated the unobserved wavelength regions with straight
%lines.

The brightness of ISD emission
is correlated with the distribution of the hydrogen column density. 
The interstellar neutral hydrogen column density of the Lockman
Hole is 4$-$6$\times10^{23}$ atoms m$^{-2}$ and that of the ecliptic plane
is 4.0$\times10^{24}$ atoms m$^{-2}$.  
To evaluate the dust thermal emission 
associated with the neutral hydrogen columns, we use the brightness
per hydrogen atom at 154 $\mu$m obtained by a rocket-borne telescope,
 $\nu I_\nu (154)/N(HI) = 3.2\times 10^{-32}$ 
W sr$^{-1}$,
the dust temperature of 16.4 K, and the gray-body dust emissivity law
proportional to $\nu^{2}$\cite{kawada}.

In the calculations below, we use the natural thermal background
for the Lockman Hole and note here that the limiting flux is higher
by about a factor of 2 for the ecliptic plane when the detection
is limited by the natural thermal background.
 
\subsection{Separated spacecraft interferometer like TPF}

For TPF, a four element array
with 3.5 m diameter apertures is considered.
The telescope apertures are cooled by radiation to 40 K.
Here we assume the emissivity of the apertures to be 5\%.
Varying  the length of the longest baseline $B$, we calculate
the SNRs of the $^nC_2$ interferometer for $n = 4$,
$D$ = 3.5 m, throughput $T = 0.1$,  $\sigma = 2 e^{-}$,
integration time $t = 100$ seconds, 
and $\Delta\nu/\nu = 0.1$.
The read noise of $\sigma = 2 e^{-}$ may appear small,
but such a low level after multiple sampling
is set as a goal for NGST\cite{ngst}.
The coherent integration time, $t$, is somewhat unknown,
but $t$ much longer than 100 seconds
is unlikely due to disturbances on the spacecrafts.

In Fig. 4, 5 $\sigma$ detection limits 
of  the $^nC_2$ (solid line) and $^nC_n^\prime$ (dashed line)
interferometers
for 3600 second
observations are given for  baseline lengths,
100 m and 1 km.
Read noise is significant at $\lambda < 5$ $\mu$m.
In the read noise limit, the detection limits of
the $^nC_2$ interferometer with a 100 m baseline and
the $^nC_n^\prime$ interferometer with a 1 km baseline 
are almost the same.
Around 10 $\mu$m, the detection limits are set by
the natural thermal background and
at $\lambda > 25$ $\mu$m, they are set by
the telescope thermal background. 
For the thermal background limit,
the detection limits are baseline independent,
because the thermal background is independent of the number of
detector pixels. 
It is noted that there will be no far-infrared capability
of TPF itself according to the current concept.

%\placefigure{fig4}

\subsection{Separated spacecraft interferometer like Darwin}

For Darwin, a six element array with 1.5 m diameter apertures
is considered. Radiation cooling is expected to cool
the telescope apertures to 30 K. Here we again assume
the emissivity of the apertures to be 5\%.
We calculate the SNRs of the $^nC_2$ and $^nC_n^\prime$
interferometers for
$n = 6$, $D =$ 1.5 m, $T = 0.1$, $\sigma =  2 e^{-}$,
integration time  $t = 100$ seconds, and 
$\Delta\nu/\nu = 0.1$.
In Fig. 5, 5 $\sigma$ detection limits for 3600 second
observations are given as
functions of wavelength for  baseline lengths,
100 m, and 1 km. For 1 km baseline,
the effect of read noise is visible up to 10 $\mu$m.
With the number of apertures, $n = 6$,
the $^nC_n^\prime$ interferometer with 1 km baseline is more
sensitive than the $^nC_2$ interferometer with 100 m baseline
in the read noise limit.

%\placefigure{fig5}

\subsection{Single spacecraft interferometer}

Here we consider a deployable interferometer with five
1.5 m apertures and the two dimensional $^nC_n$ beam combination geometry.
We assume actively cooled telescope apertures of 5 K
and emissivity of 5\%.
The system throughput $T$ is assumed to be 0.1.
We also assume  $\sigma =  2 e^{-}$,
integration time  $t = 100$ seconds, and 
$\Delta\nu/\nu = D/B$.
We consider the baseline lengths of 10, 30, and 100 m.
In Fig. 6, 5 $\sigma$ detection limits for 3600 second
observations are given as
functions of wavelength. 
The baseline length of 100 m
may be too large for the $^nC_n$ geometry
due to the narrow fractional bandwidth
and the effect of read noise. The two dimensional
$^nC_n$ interferometer has
an advantage for a compact array as discussed above.

%\placefigure{fig6}

\section{Capability for a deep far-infrared survey
 and comparison with observatories}

In this section, we investigate the potential of a TPF-like space
interferometer for a deep far-infrared survey which is 
important in studying the galaxy formation history.

We first summarize the current status of the subject.
In Fig. 7, source confusion limits 
at 90   and 170 $\mu$m of different observatories in space
predicted by the model E of Guiderdoni et al.\cite{guider98}
are compared with those estimated from the recent Infrared
Space Observatory (ISO) 
observations\cite{puget99}$^,$\cite{maruma00}. 
The model E of Guiderdoni et al. is known as one of 
the very successful models which can explain the 
Infrared Astronomical Satellite (IRAS) 60 $\mu$m
counts, ISO 15 $\mu$m counts, and cosmic far-infrared background
radiation recently detected by the Far Infrared Absolute
Spectrophotometer (FIRAS) on board COBE.

At 90 $\mu$m, 
the true confusion limit is one order of magnitude greater
than the limit estimated by the model at an angular resolution
of 10$^{\prime\prime}$ or for a telescope diameter of 2.2 m.
The model predictions of the confusion limits
of SPICA or FIRST at both 90 and 170 $\mu$m
should be regarded as lower limits.

%\placefigure{fig7}

In Fig. 8, 5 $\sigma$ detection limits of a TPF-like 
$^nC_2$ interferometer
(4 x 3.5 m) for telescope temperatures, 5, 10, 20, and 40 K with
5\% emissivity
are compared with those of NGST, SPICA, and FIRST (Nakagawa et
al. 1998). The assumed total observing time is 3600 seconds.
As for the interferometer, we assume $T = 0.1$, $\sigma = 2e^{-}$,
$\Delta\nu/\nu = 0.1$, coherent integration time
$t$ = 100 seconds, and $B$ = 100 m. 
In Fig. 9, 5 $\sigma$ detection limits of a TPF-like $^nC_n^\prime$ 
interferometer are calculated for the same conditions.

%\placefigure{fig8}

%\placefigure{fig9}

Before interpreting Figs. 8 and 9, the limitations of
the comparison between the SNRs of an interferometer and
a single telescope must be noted. For the SNR of the interferometer,
we use that for a point source. In reality
there may be multiple sources and some of them may be resolved,
in which case the complex visibility $\gamma \exp(i\phi)$ must
be obtained by (4) and be inserted into the SNR formula. 
The sensitivity of the observatory is also given for a point source.
However a single telescope can observe multiple point sources 
at once, while the interferometer needs to cover sufficient $uv$
points or baseline configurations
to recover multiple sources. So necessary observing time is likely
to be greater for the interferometer.
We proceed with the comparison with this limitation in mind.

In the near infrared, the $^nC_n^\prime$ interferometer is
more sensitive than  the $^nC_2$ interferometer.
However the spatial resolution of NGST is 
sufficient and its sensitivity is superior to both 
interferometers
for the purpose of galaxy count.
In the mid infrared, the sensitivities of the interferometers
are comparable to the sensitivity of SPICA though spatial
resolutions are different.

The advantage of the interferometers is obvious in the far infrared.
In the present concept of TPF, a telescope temperature of 40 K is assumed.
Even with 40 K telescopes, the interferometers are more sensitive
than SPICA or FIRST at $\lambda >$ 100 $\mu$m. 
It is clear from the the result of calculation
 for 5 K telescopes that an actively
cooled interferometer 
in the far infrared regardless of the beam combination geometry
is ideal for detecting faint galaxies which are
otherwise undetected due to source confusion.
Bright starburst galaxies will be
detectable at 170 $\mu$m out to z = 4 with the 
actively cooled interferometer.
The survey may not only identify 
far-infrared luminous galaxies at high redshift
as IRAS did at low redshift, but may even detect a galaxy population
which is undetectable by NGST in the near infrared due to the weakness
of stellar emission extincted by dust.

%\placefigure{fig10}

\section{Conclusions}

In this paper, we have studied 
the sensitivities of space borne infrared
interferometers for the purpose of general synthesis imaging.
Our primary conclusions follow.

(1)  We have derived the expressions for the SNRs of 
$^nC_2$ and $^nC_n$ interferometers in the presence of
source shot noise, thermal background shot noise, and detector
read noise.

(2) We have investigated
the tradeoff between the
$^nC_2$ and $^nC_n$ interferometers for a two dimensional baseline
configuration and found that the $^nC_2$ geometry is
suitable for a long baseline interferometer, while 
the  $^nC_n$ geometry is fitted to a short baseline deployable interferometer.
For a one dimensional baseline configuration, 
we have found that the $^nC_n^\prime$ interferometer
is superior especially in the case of read noise limited observations.

(3)
We have presented the detection limits of separated spacecraft
interferometers like TPF and Darwin in the presence of
the natural thermal background as functions of wavelength.
The comparison of a TPF-like interferometer with
NGST, SPICA and FIRST has revealed that at $\lambda > 100$ $\mu$m,
an interferometer  is more
sensitive than SPICA or FIRST.
This is true with only radiation cooling of the apertures.

We thank David Saint-Jacques and the anonymous referee
for comments on the manuscript.
TN is supported by Grant-in-Aid for Scientific Research
of the Japanese Ministry of Education, Culture, Sports,
and Science (No. 10640239).

\appendix

\section{Variance $V[i_2(q)]$}

The derivation of the variance $V[i_2(q)]$ is analogous
to that of $V[i_1(q)]$ but more complicated. We utilize
the results by PK89 whenever possible.

\begin{eqnarray}
V[i_2(q)] & = & <i_2 (q)^2> - <i_2(q)>^2 \nonumber \\ 
 & = & \sum_{i<j}\sum_{k<l}[<Re(y_{ij})Re(y_{kl})>-
<Re(y_{ij})><Re(y_{kl})>]\cos(\omega_{ij}q)\cos(\omega_{kl}q)
\nonumber \\
 & & -[<Re(y_{ij})Im(y_{kl})>-<Re(y_{ij})><Im(y_{kl})>]
 \cos(\omega_{ij}q)\sin(\omega_{kl}q) \nonumber \\
 & & -[<Im(y_{ij})Re(y_{kl})>-<Im(y_{ij})><Re(y_{kl})>]
 \sin(\omega_{ij}q)\cos(\omega_{kl}q) \nonumber \\
 & & +[<Im(y_{ij})Im(y_{kl})>-<Im(y_{ij})><Im(y_{kl})>]
 \sin(\omega_{ij}q)\sin(\omega_{kl}q) 
\end{eqnarray}

As in \S2,
the problem of estimating the variance in the image thus
reduces to that of estimating three types of covariance
term: cov$[Re(y_{ij}),Re(y_{kl})]$,
cov$[Re(y_{ij}),Im(y_{kl})]$,
cov$[Im(y_{ij}),Im(y_{kl})]$.
Using (20),  we obtain the covariance
of the real components of a pair of fringe phasors as

\begin{eqnarray}
cov[Re(y_{ij}),Re(y_{kl})] & = &
\sum_p (<k(p)> + \sigma^2) \cos(p\omega_{ij})\cos(p\omega_{kl}) 
\nonumber \\
& = & \sum_p \Bigl[<Q_0>n + \sigma^2
+ <Q_0>\sum_{g<h} \gamma_{gh}\cos(p\omega_{gh}+\phi_{gh})\Bigr]
\nonumber \\
& & \ \ \ \ \times \cos(p\omega_{ij}) \cos(p\omega_{kl})
\end{eqnarray}

Every cosine can be written as the sum of exponential functions:

\begin{eqnarray}
\sum_p \cos(p\omega_{ij}) \cos(p\omega_{kl})
& = & \sum_p \frac{\exp(ip\omega_{ij})+\exp(-ip\omega_{ij})}{2}
             \frac{\exp(ip\omega_{kl})+\exp(-ip\omega_{kl})}{2}
\nonumber \\
& = & \sum_p \Bigl[\frac{\exp(ip(\omega_{ij}+\omega_{kl}))}{4}            
             +\frac{\exp(ip(\omega_{ij}-\omega_{kl}))}{4} \nonumber \\
& &          +\frac{\exp(-ip(\omega_{ij}+\omega_{kl}))}{4}
             +\frac{\exp(-ip(\omega_{ij}-\omega_{kl}))}{4}\Bigr]
\nonumber \\
& = & P \frac{\delta_{ik} \delta_{jl}}{2}.
\end{eqnarray}

In (57), from nonredundancy of baselines, 
$\pm\omega_{ij}\pm\omega_{kl}\neq 0$ for $i \neq k, j \neq l$.
Similarly to the two frequency case, the three frequency case
can be computed.

Following PK89, we now impose an additional
condition to ordinary nonredundancy in order to simplify
the calculations. This condition, hereafter referred to 
as the nonredundancy of triangles, concerns three-frequency
combinations. Specifically, we assume that

\begin{equation}
\omega_{gh} \pm \omega_{ij} \pm \omega_{kl} \neq 0,
\end{equation}

unless (gh), (ij), and (kl) form the sides of a triangle.
This condition is easily fulfilled for a space interferometer
with a small number of apertures. With the nonredundancy 
of triangles for $i<j$ and $k < l$,

\begin{equation}
\sum_p \sum_{g<h} \gamma_{gh}
\cos(p\omega_{gh}+\phi_{gh})\cos(p\omega_{ij}) \cos(p\omega_{kl})
= \frac{\gamma\cos\phi}{2}\Delta_{ij,kl}
\end{equation}

where the symbol $\Delta_{ij,kl}$ unless ${ij}$ and ${kl}$
baselines form two sides of a triangle, in which case it equals
1, and $\gamma \exp(i\phi)$ is the phasor on the third side
of that triangle.

From Eqs (57) and (59),

\begin{equation}
{\rm cov}[Re(y_{ij}),Re(y_{kl})]
= \frac{<M>n+P\sigma^2}{2}\delta_{ik}\delta_{jl}
+ <M> \frac{\gamma \cos\phi}{2} \Delta_{ij,kl}.
\end{equation}

Similarly,
 
\begin{equation}
{\rm cov}[Im(y_{ij}),Im(y_{kl})]
= \frac{<M>n + P\sigma^2}{2}\delta_{ik}\delta_{jl}
\mp <M> \frac{\gamma \cos\phi}{2} \Delta_{ij,kl}.
\end{equation}

Everywhere in this section, the upper sign (or expression)
is the correct one when the sides $(ij)$ and $(kl)$
of the triangle meet at that vertex for which the
label has a value intermediate to those of the two
vertices, i.e., either when $i<j=k<l$ or $k < l=i< j$.
The lower sign (or expression) is the correct one otherwise,
i.e., when $i=k$ or when $j=l$.

Evaluation of ${\rm cov}[Re(y_{ij}),Im(y_{kl})]$ is 
identical with that for the case with Poisson noise alone
as obtained in PK89:

\begin{equation}
{\rm cov}[Re(y_{ij}),Im(y_{kl})]
= <M> \frac{\gamma \sin\phi}{2}\Delta_{ij,kl}
\times \left\{ \begin{array}{c}
                    +1 \\ 
             sgn(i-j) \mbox{ for } i=k, sgn(i-k) \mbox{ for } j=l
             \end{array}
       \right \}, 
\end{equation}

where $sgn(x)$ is $+1$ if $x>0$ and $-1$ if $x<0$.  

Furthermore,

\begin{equation}
{\rm cov}[Re(y_{ij}),Im(y_{kl})] = \pm {\rm
cov}[Im(y_{ij}),Re(y_{kl})].
\end{equation}

Now we evaluate the variance,

\begin{eqnarray}
V[i_2(q)] & = & \frac{n<M>+P\sigma^2}{2}\sum_{i<j}
[\cos^2(q\omega_{ij}) + \sin^2(q\omega_{ij})] \nonumber \\
& & 
+ \frac{<M>}{2}\Bigl\{  \sum_{i<j}\sum_{k<l}\Delta_{ij,kl}
\cos[q(\omega_{ij}\pm\omega_{kl})]\gamma\cos\phi 
\nonumber \\
& & -\sum_{i<j=k<l}\sin[q(\omega_{ij}+\omega_{kl})]\gamma_{il}\sin\phi_{il}
\nonumber \\
& & -\sum_{k<l=i<j}\sin[q(\omega_{ij}+\omega_{kl})]\gamma_{kj}\sin\phi_{kj}
\nonumber \\
& &-\sum_{j<l}(n-l+j-1)\sin[q(\omega_{il}+\omega_{ij})]\gamma_{jl}\sin\phi_{jl}
\nonumber \\
& &-\sum_{l<j}(n-j+l-1)\sin[q(-\omega_{il}+\omega_{ij})]\gamma_{lj}\sin\phi_{lj}.
\end{eqnarray}

By relabeling the indices slightly and combining the various
sums using simple trigonometric identities, we obtain the 
final expression of $V[i_2(q)]$, (34).

\section{Remapping of Pupils}

The anonymous referee pointed out the possibility of
pupil remapping which changes the interpretation of
the interferometer types. We do not consider this 
possibility in the main text, but mention here the view
based on it.

The distinction between the $^nC_2$, $^nC_n$, and $^nC_2^\prime$
refers actually to the ``output pupil'' (arrangement of the beams
immediately before fringe detection) rather than to the 
configuration of the interferometer as a whole. One could use
a two dimensionally distributed array with a linear nonredundant
arrangement of the beams before the detector like the one shown in Fig. 3,
with a remapping stage during data processing. Any two dimensional
interferometer can therefore accommodate spectral dispersion in principle.

However such a scheme would require additional optics, affecting
the system throughput and thermal environment. It also makes 
the image reconstruction procedure complex. We therefore would not
include this scheme in this comparative analysis.

%\begin{thebibliography}{}
%\bibitem{paper1}  Author names here, ``title,''  
%Journal name {\bf volume number bold}, inclusive pages (year).
%\bibitem{Trigt97}  C. van Trigt, ``Visual system-response
% functions and estimating  reflectance,'' \josaa
% {\bf 14}, 741-755 (1997).
%\end{thebibliography}

%% Figures and tables after References
\newpage

\noindent Fig. 1.
 Focal plane interferometer and a detector
array.  This configuration exists for each baseline of 
an $^nC_2$ interferometer. The effect of shot noise 
on photodetection process has been fully formulated for this
situation. 
For spectrally broad light, the fringe can be dispersed
in the cross fringe direction if  a two-dimensional 
detector is used.

\vskip 5mm
\noindent Fig. 2.
Two-dimensional $^3C_3$ interferometer.
Fringes are superposed at the focal plane of a two dimensional
detector. Since the fringes cannot be dispersed, the spectral
bandwidth of this interferometer is limited.

\vskip 5mm
\noindent Fig. 3.
One-dimensional $^3C_3^\prime$ interferometer.
One dimensional fringes
 are dispersed in the cross
fringe direction and superposed at the focal plane of
a two dimensional detector. Spectrally broad light can be observed
by an  $^nC_n$ interferometer with
one dimensional baselines.

\vskip 5mm
\noindent Fig. 4.
Detection limits of a TPF-like
interferometer.
5 $\sigma$ detection limits are plotted as functions
of wavelength for the maximum baseline lengths given in
the figure for an $^nC_2$ interferometer (solid)
and a one-dimensional $^nC_n^\prime$ interferometer (dash). 
Assumptions about other parameters are given
in the text.

\vskip 5mm
\noindent Fig. 5.
Detection limits of a Darwin-like
interferometer. 
5 $\sigma$ detection limits are plotted as functions
of wavelength for the maximum baseline lengths given in
the figure for an $^nC_2$ interferometer (solid)
and a one dimensional $^nC_n^\prime$ interferometer (dash). 
Assumptions about other parameters are given
in the text.

\vskip 5mm
\noindent Fig. 6.
Detection limits of a five element compact array with
artificially cooled apertures.  
5 $\sigma$ detection limits of a two-dimensional $^nC_n$ 
interferometer  are plotted as functions
of wavelength for the maximum baseline lengths given in
the figure. Assumptions about other parameters are given
in the text.

\vskip 5mm
\noindent Fig. 7.
Confusion limits of monolithic observatories.
The confusion limits predicted from the model E of
Guiderdoni et al. \cite{guider98} (solid line) are compared with
those estimated from ISO observations 
by Matsuhara et al.\cite{maruma00} (polygon) and 
Puget et al.\cite{puget99} (triangle).

\vskip 5mm
\noindent Fig. 8.
Comparison with monolithic observatories.
The sensitivity of an $^nC_2$ interferometer
composed of four 3.5 m apertures
is plotted for telescope temperatures of
5, 10, 20, and 40 K with emissivity
of 5\%. The maximum baseline length is assumed to
be 100 m. 
Dashed lines denote the sensitivities
of NGST, SPICA, and FIRST given  for comparison.

\vskip 5mm
\noindent Fig. 9.
Comparison with monolithic observatories.
The sensitivity of a one dimensional $^nC_n^\prime$ interferometer
composed of four 3.5 m apertures
is plotted for telescope temperatures of
5, 10, 20, and 40 K with an emissivity
of 5\%. The maximum baseline length is assumed to
be 100 m. 
Dashed lines denote the sensitivities
of NGST, SPICA, and FIRST given  for comparison.

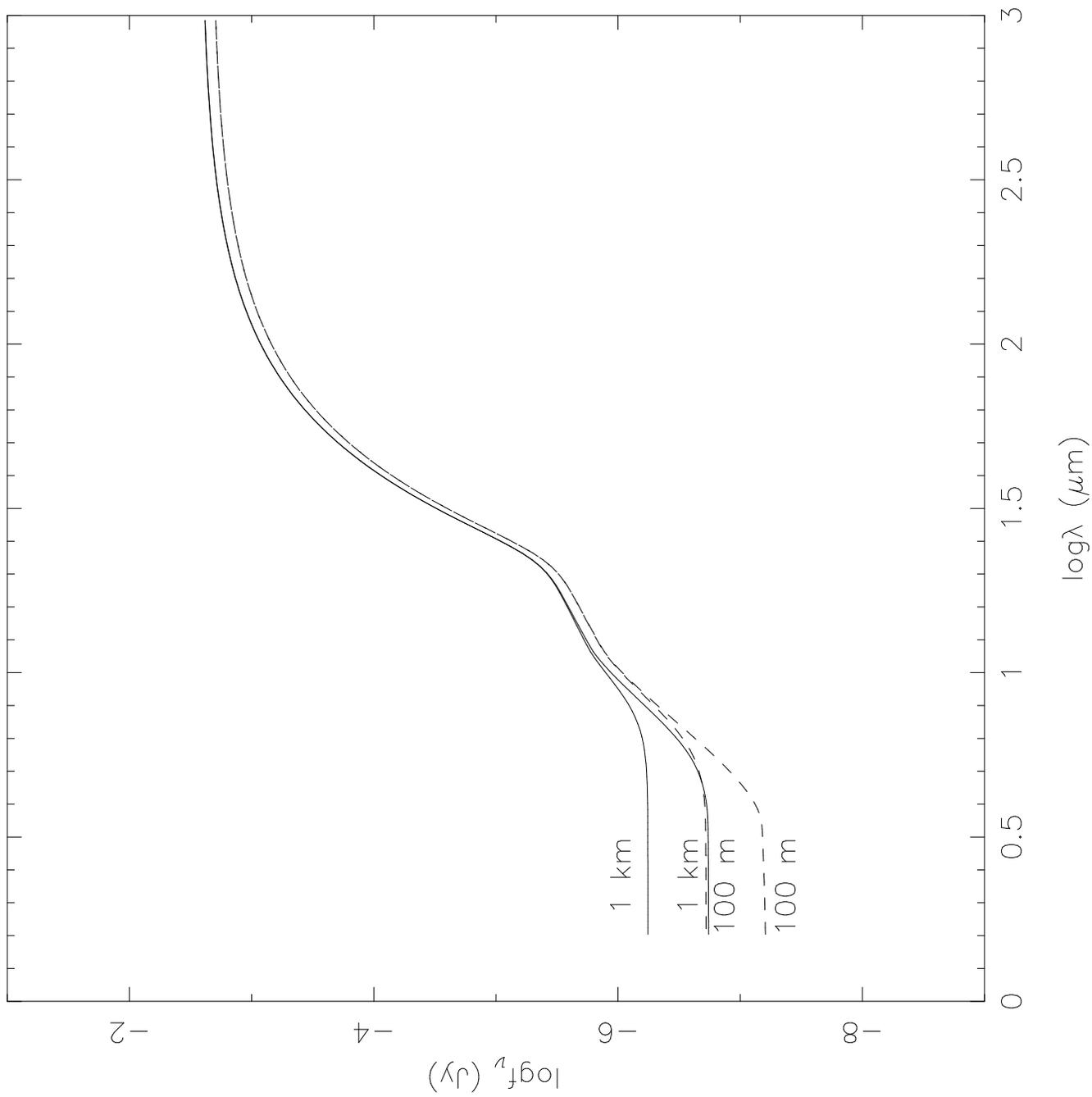
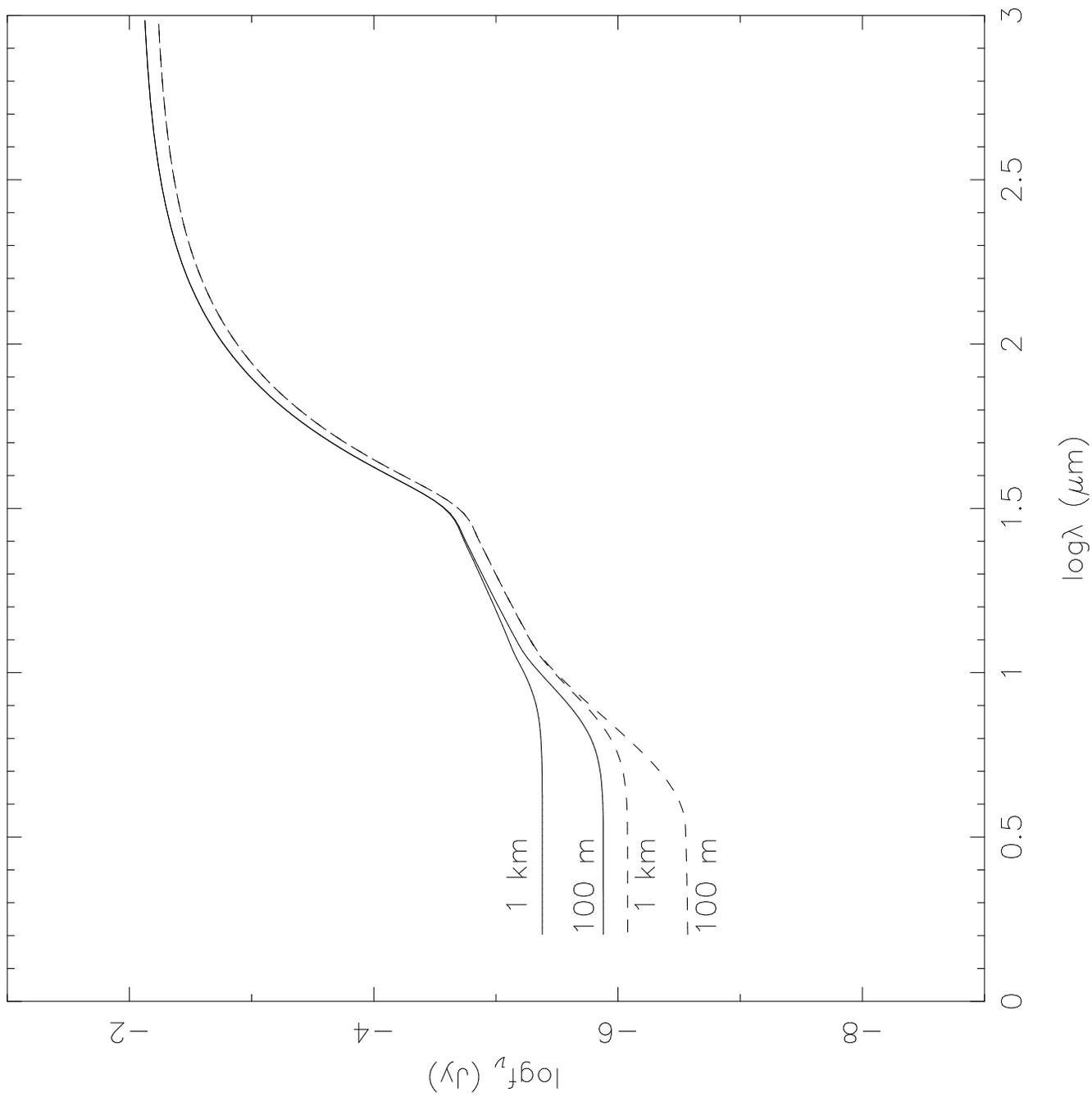
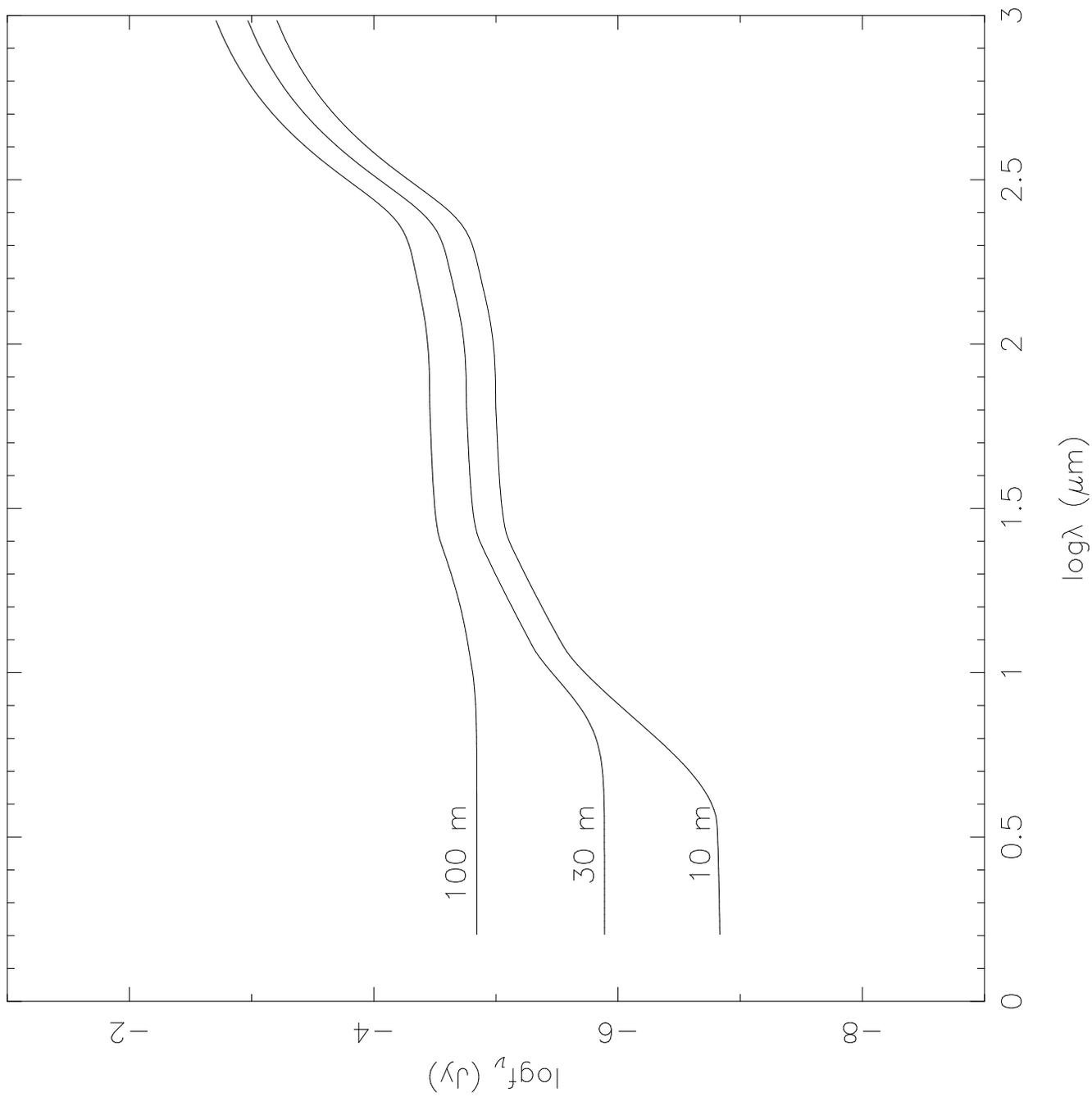
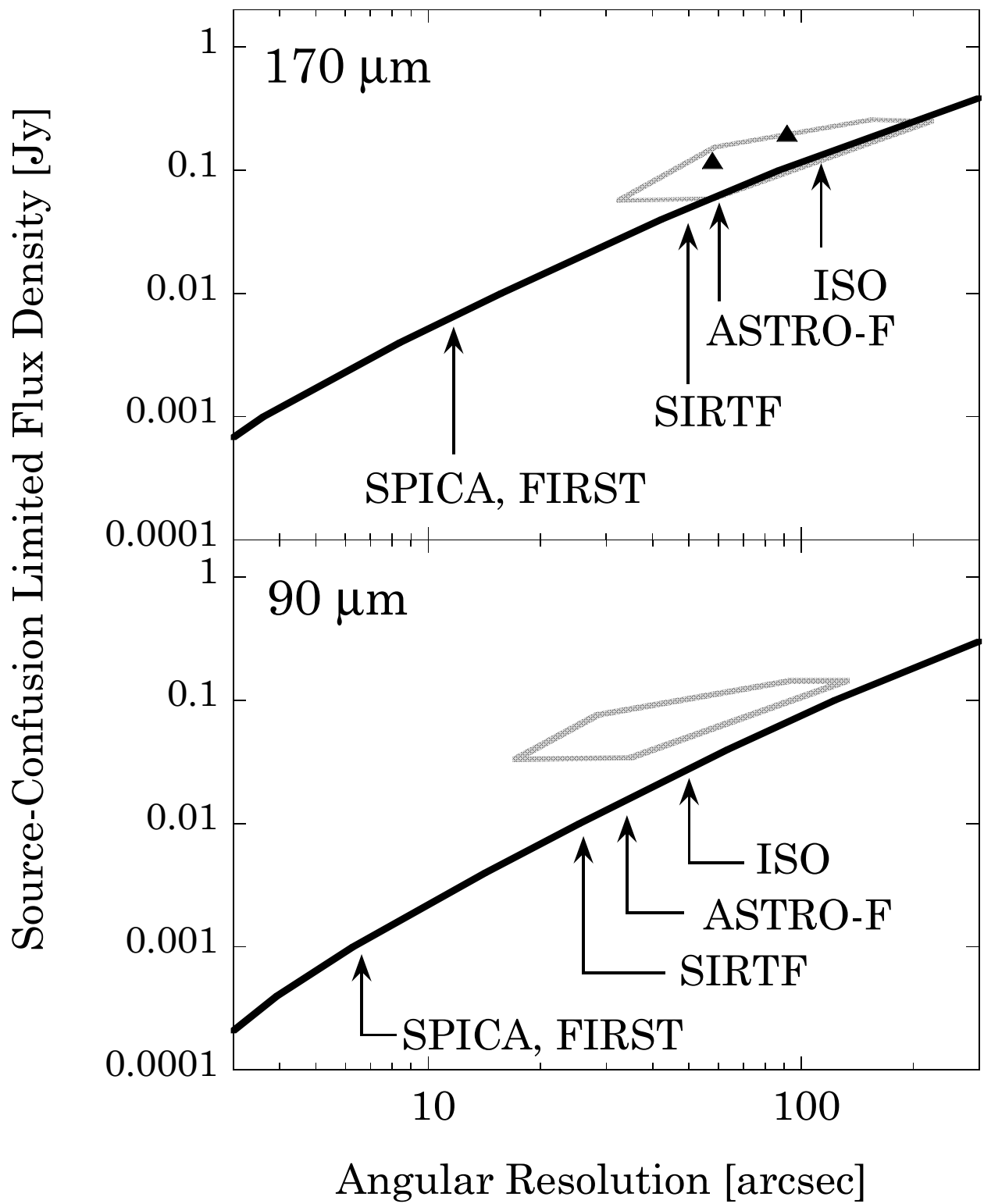
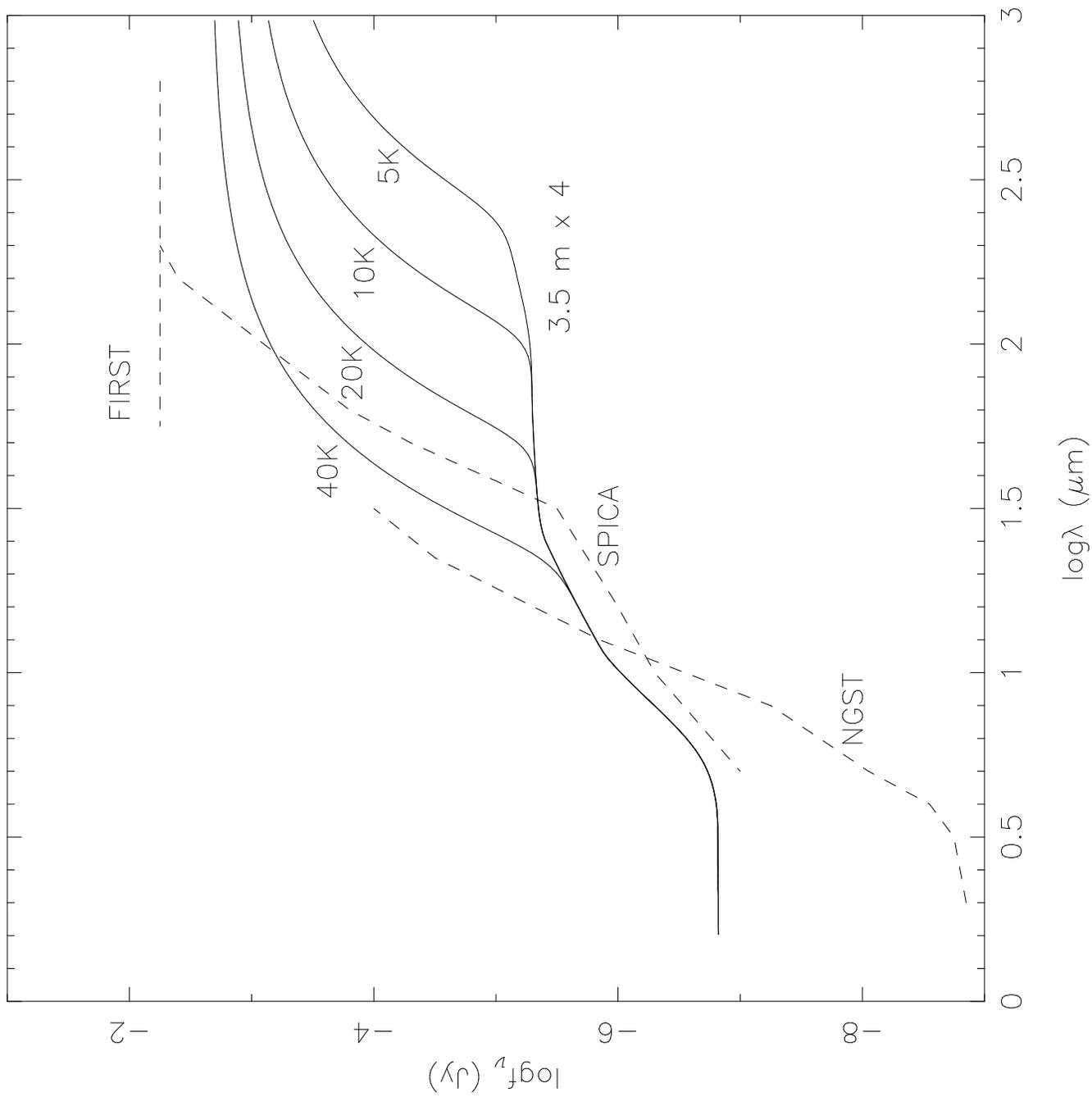
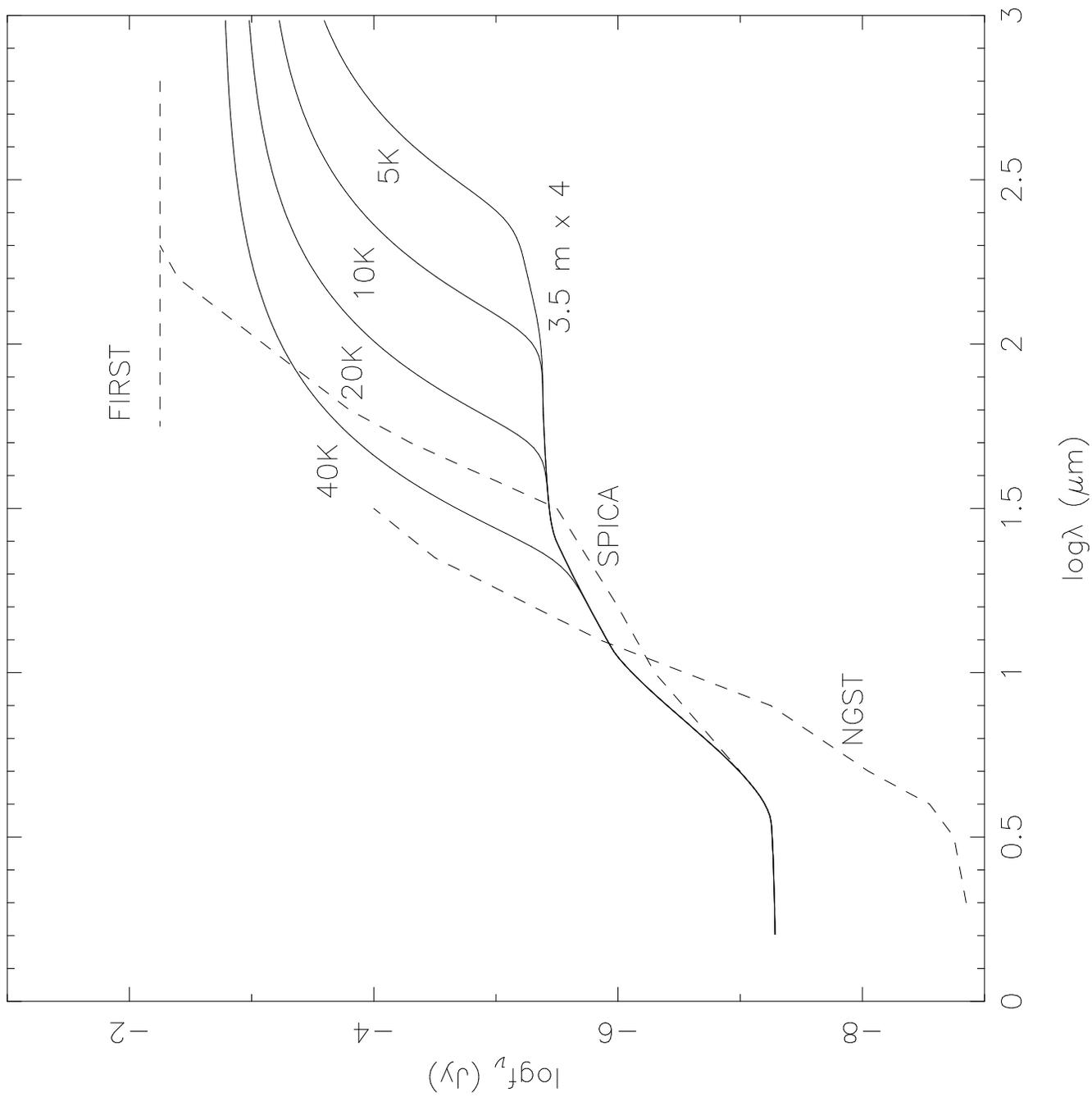
\end{document}